\documentclass[aps,prl,amsmath,twocolumn,superscriptaddress,showpacs]{revtex4}

\usepackage{graphicx}

\begin{document}

\title{Spatially resolved manipulation of single electrons\\ in quantum dots using a scanned probe}

\author{A. Pioda}
\affiliation{Solid State Physics, ETH Z\"urich, 8093 Z\"urich, Switzerland}

\author{S. Ki\v{c}in}
\affiliation{Solid State Physics, ETH Z\"urich, 8093 Z\"urich, Switzerland}

\author{T. Ihn}
\affiliation{Solid State Physics, ETH Z\"urich, 8093 Z\"urich, Switzerland}

\author{M. Sigrist}
\affiliation{Solid State Physics, ETH Z\"urich, 8093 Z\"urich, Switzerland}

\author{A. Fuhrer}
\affiliation{Solid State Physics, ETH Z\"urich, 8093 Z\"urich, Switzerland}

\author{K. Ensslin}
\affiliation{Solid State Physics, ETH Z\"urich, 8093 Z\"urich, Switzerland}

\author{A. Weichselbaum}
\affiliation{Department of Physics and Astronomy, Ohio University, Athens, Ohio 45701-2979}

\author{S. E. Ulloa}
\affiliation{Department of Physics and Astronomy, Ohio University, Athens, Ohio 45701-2979}

\author{M. Reinwald}
\affiliation{Institut f\"ur experimentelle und angewandte Physik, Universit\"at Regensburg, Germany}

\author{W. Wegscheider}
\affiliation{Institut f\"ur experimentelle und angewandte Physik, Universit\"at Regensburg, Germany}

\date{\today}

\begin{abstract}
The scanning metallic tip of a scanning force microscope was coupled capacitively to electrons confined in a lithographically defined gate-tunable quantum dot at a temperature of 300~mK. Single electrons were made to hop on or off the dot by moving the tip or by changing the tip bias voltage owing to the Coulomb-blockade effect. Spatial images of conductance resonances map the interaction potential between the tip and individual electronic quantum dot states. Under certain conditions this interaction is found to contain a tip-voltage induced and a tip-voltage independent contribution.
\end{abstract}

\pacs{73.21.La, 73.63.Kv, 73.23.Hk}
\maketitle

Single-electron charging and tunneling are two fundamental phenomena essential in many nanoscale systems and materials ranging from man-made nanodevices down to single molecules and atoms. Semiconductor quantum dots offer the opportunity to study these quantum effects in a well controllable way \cite{Grabert92,Kouwenhoven97,Reimann02}, and they have been suggested as building blocks within quantum information processing schemes \cite{Loss98}. They allow the manipulation of single electron charges \cite{Grabert92}, spins \cite{Tarucha00,Luscher01,Rokhinson01,Fuhrer03} and orbital quantum states \cite{Fuhrer01}. Scanning probe techniques give local microscopic access to nanostructures. For example, using such methods branched electron flow and quantum interference past a quantum point contact were demonstrated \cite{Topinka01,Topinka03}. Conductance imaging within a point contact \cite{Crook00} and a quantum billiard \cite{Crook03} were reported. Single-charge sensitivity was reported using scanning single-electron transistors \cite{Yoo97,Yacoby99,Zhitenev00,Ilani04} or the subsurface charge accumulation technique \cite{Finkelstein00}.  Local spectroscopy of carbon nanotube quantum dots was achieved \cite{Woodside02}. Here we show that in lithographically defined semiconductor quantum dots, single electrons on the dot can be controlled by moving the tip. Spatial images of conductance resonances lead directly to quantitative spatial maps of the interaction potential between the tip and individual electrons.

\begin{figure}[bthp]
\centering{\includegraphics[width=3.0in]{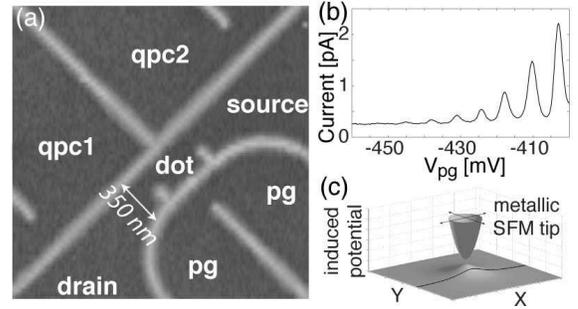}}
\caption{\label{Fig1} (a) Topography of the structure measured at room temperature. (b) Conductance resonances measured by varying $V_{\text{pg}}$ at 300~mK with 20~$\mu$V applied between source and drain. (c) Schematic illustration of the local potential induced below the tip.}
\end{figure}

The quantum dot sample has been fabricated on an AlGaAs-GaAs heterostructure containing a two-dimensional electron gas (2DEG) 34~nm below the surface, with density  $5\times 10^{11}$~cm$^{-2}$ and  mobility  $450'000$~cm$^2$/Vs at 4.2~K. The quantum dot has been defined by room temperature local anodic oxidation with a scanning force microscope (SFM) which allows to write oxide lines on a semiconductor surface that locally deplete the 2DEG underneath \cite{Fuhrer02}. Figure~\ref{Fig1}(a) is a topography image of the sample surface showing the quantum dot and three adjacent quantum point contacts that are not analyzed in the present experiment. The electron number in the quantum dot can be controlled by the lateral plunger gate pg. The gates qpc1 and qpc2 are used for tuning the coupling to source and drain. In the measurements shown below, the dot has been operated in the Coulomb-blockade regime by applying appropriate negative voltages to these gates. Figure~\ref{Fig1}(b) shows conductance resonances as a function of the voltage $V_{\text{pg}}$. The average charging energy is about~1.4 meV as determined from Coulomb-blockade diamonds. We estimate the spacing of single-particle quantum states from the geometry of the dot to be about 30~$\mu$eV.

Experiments are performed with a SFM in a $^3$He cryostat with a base temperature of 300~mK. The scanning sensor is an electrochemically sharpened PtIr tip (radius a few tens of nm) mounted on a piezoelectric tuning fork. Details of this low-temperature SFM setup can be found in Ref. \onlinecite{Ihn04}. Figure~\ref{Fig1}(c) shows schematically the local potential induced in the 2DEG by the tip. Its strength and sign depends on the tip bias voltage, its geometric shape and extent reflects the tip shape and the tip-sample separation. Moving the tip in the vicinity of the quantum dot affects its conductance via the induced potential by changing the coupling strength to source and drain and by shifting the energy levels. Scanning gate images are maps of the dot's conductance as a function of tip position. Electron flow between the tip and the electron gas is suppressed due to the vacuum gap between tip and surface and the 34~nm insulating barrier between surface and 2DEG. Experiments have been performed on different samples and for different cooldowns of the same sample. Here we present one concise set of data from one experiment.

\begin{figure}[tbhp]
\centering{\includegraphics[width=3.0in]{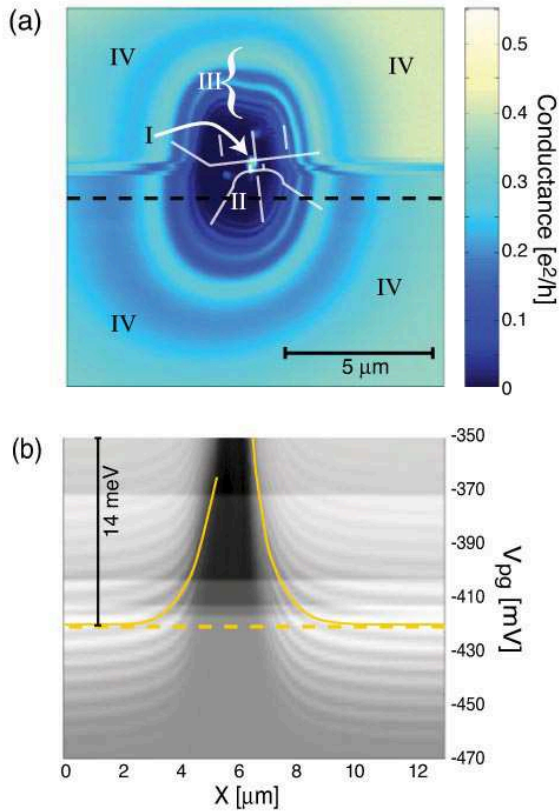}}
\caption{\label{Fig2} (Color online) (a) Scanning gate image measured with source--drain bias $V_{\text{SD}}=20\,\mu$V and $V_{\text{pg}}=-420$~mV and the tip ($V_{\text{tip}}=0$) scanned in feedback at a distance of a few nanometers from the surface. The oxide lines defining the structure have been drawn as white lines. Regions (I) to (IV) are described in the text. (b) Dot current obtained when scanning the tip in feedback along the horizontal dashed line in (a), for different $V_{\text{pg}}$. The horizontal dashed line in (b) corresponds to $V_{\text{pg}}$ used for the scan in (a).}
\end{figure}

Figure~\ref{Fig2}(a) shows a scanning-gate image taken at a gate voltage tuning the dot, in the absence of the tip, close to the Coulomb blockade regime. Four different regions labeled I to IV may be distinguished: (I) The central bright peak of increased conductance when the tip is directly above the quantum dot, (II) the dark blockaded region where no measurable current flows through the dot, (III) the region of concentric conductance peaks around the dark blockaded region and (IV) a region of weakly varying but relatively high conductance at larger distance from the dot.
Considering a tip-induced potential like that shown in Fig.~\ref{Fig1}(c), the phenomenology in regions (II) to (IV) can be interpreted in a straightforward way: The constant conductance region (IV) arises when the tip is too far away from the dot to have strong influence on its conductance. The dot is relatively open and the conductance is rather high. When the tip comes closer [region (III)], the potential energy of electrons in the dot is increased and they spill over into source or drain one by one \cite{Woodside02}. At the same time the point contacts become sufficiently pinched to observe clear Coulomb blockade. Each of the conductance peaks forming the ring-like pattern indicates that a single electron leaves the quantum dot. These conductance oscillations correspond to the oscillations displayed in Fig.~\ref{Fig1}(b) where they are induced by the plunger gate rather than by the moving tip. The closed curve around the dot defined by the position of a particular conductance peak corresponds to a constant potential in the quantum dot. Therefore conductance peaks map out contours of the interaction potential between tip and quantum dot electrons. The non-circular shape of the contour lines (independent of the scan direction) suggests that the tip is not symmetric with respect to rotation around the z-axis normal to the sample surface. This peculiarity in the tip shape is due to the electrochemical etching process used to sharpen the tip before use. When the tip is scanned even closer to the dot in region (II), the current through the dot becomes smaller than our current resolution (
50~fA) because the tip-induced potential pinches off the source--drain coupling. Numerical simulations show that this pinch-off occurs before the dot itself is completely depleted \cite{Weichselbaum03}. The enhanced conductance in the central region (I) is unexpected. The tip-induced potential shown in Fig.~\ref{Fig1}(c) is most repulsive there for electrons. This conductance enhancement effect depends on cooldown. In Fig.~\ref{Fig2}(a) a horizontal stripe of modified conductance occurs when the tip crosses the dot. Within this stripe, conductance peaks are shifted outwards reproducibly over successive scans. The direction of the stripe rotates with the scan direction. We attribute this feature to tip-induced charge rearrangements in the vicinity of the dot.

We map the interaction potential between the tip and the dot quantitatively using the conductance peaks in region (III). The tip is scanned with $V_{\text{tip}}=0$~V along a line in x-direction offset from the dot center by 1.3~$\mu$m and running through regions (II) to (IV) [horizontal dashed line in Fig.~\ref{Fig2}(a)]. Such scans are repeated for a series of increasing plunger gate voltages. The results are summarized in Fig.~\ref{Fig2}(b). The plunger gate voltage axis can be converted to energy using the lever arm \cite{Kouwenhoven97}. Any line following a particular conductance peak in this plot quantitatively represents the tip-induced potential along a similar line as indicated in Fig.~\ref{Fig1}(c). In Fig.~\ref{Fig2}(b) the tip-induced potential maximum at a distance of 1.3~$\mu$m from the tip apex is more than 14~meV, comparable to the Fermi energy. We conclude that the electron gas is depleted directly under the tip at $V_{\text{tip}} = 0$~V.

\begin{figure}[tbhp]
\centering{\includegraphics[width=3.0in]{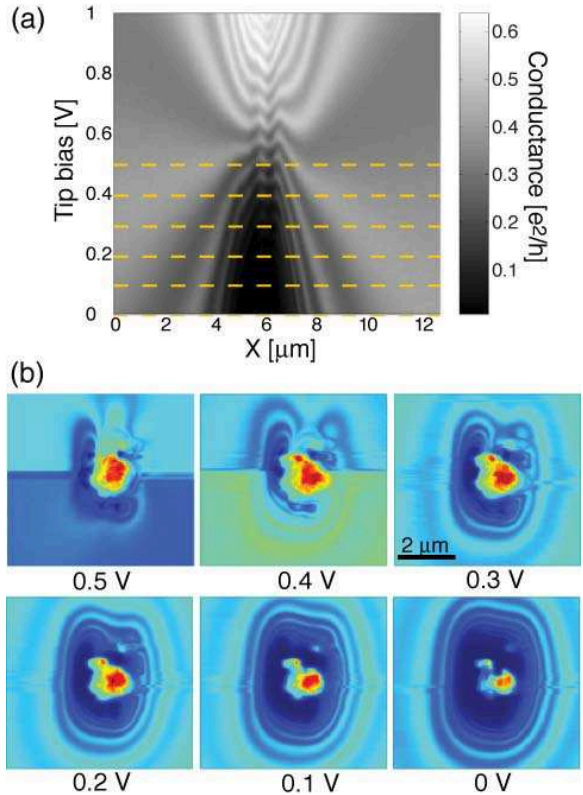}}
\caption{\label{Fig3} (Color online) (a) Conductance of the quantum dot vs. tip bias along the horizontal dashed line in Fig.~\ref{Fig2}(a) at $V_{\text{pg}}=Ð390$~mV and 200~nm tip-sample separation. (b) Scanning gate measurements for different $V_{\text{tip}}$ as indicated by the dashed lines in (a) taken at a tip--sample separation of 120~nm and $V_{\text{pg}}=-390$~mV. The scanned area is $6.5\times 6.5\,\mu$m$^2$, the conductance ranges from zero (dark) to $1.1 e^2/h$ (bright).}
\end{figure}

We proceed with minimizing the tip-induced potential by setting an appropriate tip bias voltage in analogy with the standard method established in Ref.~\onlinecite{Yoo97}. Line scans have been performed along the line indicated in Fig.~\ref{Fig2}(a) for a sequence of increasing $V_{\text{tip}}$. Figure~\ref{Fig3}(a) shows the conductance as a function of tip position and tip voltage. The evolution of individual conductance resonances differs below and above the symmetry line around $V_{\text{tip}}=+0.56$~V. Below this voltage the curvature is concave, above it is convex. The reason for this behavior is that the tip-induced potential is repulsive for $V_{\text{tip}}<0.56$~V and attractive for $V_{\text{tip}}>0.56$~V implying that $V_{\text{tip}}=0.56$~V can be called the least invasive voltage. This  offset in the effective voltage has contributions from the work function difference between the PtIr tip and the GaAs heterostructure and from the Fermi-level pinning at the GaAs surface. Taking both contributions together, the net offset voltage is expected to be in the range between 0.4 and 0.8~eV, in agreement with the observations and previous measurements \cite{Vancura03}.

Even at $V_{\text{tip}}=0.56$~V in Fig.~\ref{Fig3}(a) there is a kind of zig-zag motion of conductance peaks for $5.5\,\mu\text{m}<x<7\,\mu\text{m}$. It implies that even when the tip voltage does not induce any charge in the dot on average, there is still a local and finite tip-induced potential independent of tip bias. This latter property of the feature can be recognized in Fig.~\ref{Fig3}(a). We conclude that the total tip-induced potential may contain two contributions one of which depends on the tip voltage whereas the other is voltage-independent.

The tip-voltage dependence of the conductance is further illustrated in the conductance maps shown in Fig.~\ref{Fig3}(b) taken for different $V_{\text{tip}}$ values. With increasing $V_{\text{tip}}$, conductance resonance fringes are seen to contract towards the central dot as a result of the tip voltage dependent contribution of the tip-induced potential. The region of enhanced conductance in the center of the images [related to region (I) in Fig.~\ref{Fig2}(a)] eventually merges with the fringes. Beyond the least invasive tip voltage (not shown), the resonance fringes expand with increasing voltage as suggested by Fig.~\ref{Fig3}(a).

\begin{figure}[tbhp]
\centering{
\includegraphics[width=3.0in]{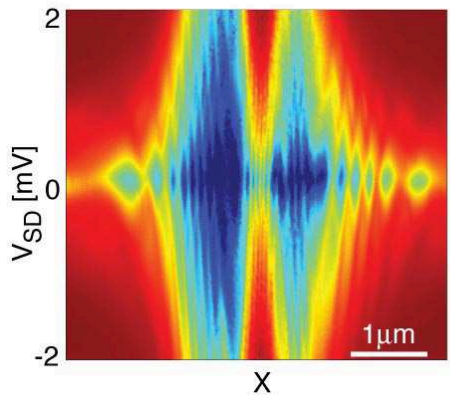}}
\caption{\label{Fig4} (Color online) Coulomb-blockade diamonds as a function of tip position along the line shown in Fig.~\ref{Fig2}(a) and the source--drain bias voltage $V_{\text{SD}}$. The tip was lifted above the surface by 150~nm and the tip voltage was 0.2~V. The color scale represents the logarithm of the dot conductance.}
\end{figure}

With the presented technique, Coulomb-blockade diamonds, i.e., regions of constant integer electron number in the plane of source--drain voltage $V_{\text{SD}}$ and plunger gate voltage $V_{\text{pg}}$  can be measured in an alternative way (Fig.~\ref{Fig4}). Here, the tip position was varied along the dashed horizontal line in Fig.~\ref{Fig2}(a) with constant $V_{\text{pg}}$. The extent of the dark diamond shaped regions varies in $x$-direction and in $V_{\text{SD}}$-direction. The slope of the diamond boundaries---in conventional $V_{\text{pg}}$-dependent measurements interpreted in terms of the lever arm of the plunger gate---represents in our case the change in energy of single-electron state ($e\Delta V_{\text{SD}}$) per change in  tip position ($\Delta x$). $\Delta V_{\text{SD}}/\Delta x=E_x$ gives the $x$-component of the tip-induced electric field in the dot. From Fig.~\ref{Fig4} we determine electric fields of the order $10^3$--$10^4$~V/m. As the tip comes closer to the dot, the electric field increases and the diamonds become more extended in $V_{\text{SD}}$-direction. At the smallest tip--dot spacing, i.e., in the center of the image, the conductance of the dot increases again. This behavior is related with region (I) in Fig.~\ref{Fig2}(a) and the zig-zag in Fig.~\ref{Fig3}(a).

The white spot in the center of the scanning gate images [Figs. 2(a) and 3 (b)]
and the zig-zag motion of peaks in Fig. 3(a) are not caused by leakage currents through the oxide lines, which are below experimental resolution.
The shape and exact position of the spot depend on the in-plane gate voltages applied, although its position always remains closely tied to the quantum dot.
A similar feature was also observed for another quantum dot structure in another experiment. However, in a second cooldown of the same sample, that lead to the application of considerably smaller plunger gate voltages, the white spot was absent.

Tip-voltage independent potentials can be created either by fixed (immobile) charges or by voltages applied to the gates. Fixed charges arise, e.g., close to the oxide lines, at the sample surface and in the doping plane of the structure. Both, fixed charges and gate voltages, lead to inhomogeneous electric fields subject to tip-position dependent screening by the metallic tip. Numerical simulations \cite{Weichselbaum04} indicate that screening of fixed charges alone is insufficient to account for the observed conductance peak in region (I). Although the exact mechanism leading to the peak in region (I) of Fig.~\ref{Fig2}(a) and to the zig-zag motion in Fig.~\ref{Fig3}(a) remains to be further investigated, screening of inhomogeneous electric fields caused by gate voltages may be a realistic scenario.
Large conductance changes are expected because the tunneling contacts to source and drain are exponentially sensitive to changes of the local potential.

The measurements discussed in this paper are related to those on carbon nanotubes \cite{Woodside02} where quantum dots form spontaneously at kinks or other imperfections between contacts. Our experiment gives local access to reproducibly fabricated laterally defined quantum dots for which the source--drain coupling and the number of electrons can be tuned. The method lends itself for a very controlled investigation of different geometries, e.g., quantum rings \cite{Fuhrer03,Fuhrer01}, or coupled quantum systems. An important difference to Refs.~\onlinecite{Topinka01,Topinka03,Crook00,Crook03,Woodside02} is the use of PtIr tips in our experiment as compared to doped Silicon cantilevers. Our tip can be regarded as a metallic electrode with electric field lines normal to the surface. In Ref.~\onlinecite{Woodside02} the experiments had to be performed with the tip more than 100~nm above the surface in order to avoid charge rearrangements in the sample. Our method of compensating the work function difference between the tip and the sample and keeping the net applied voltage small allowed stable operation of the sample over weeks with the tip scanning directly on the surface which enhances the spatial resolution. We found irreversible changes in sample properties in different experiments only at larger applied effective tip voltages.

Our scanning gate measurements on a quantum dot demonstrate that single electrons can be manipulated one by one with a macroscopic scanning tip. Advanced experiments and setups promise that it may be feasible to get a handle on the local quantum mechanical probability distributions of electronic states in quantum dot structures and coupled mesoscopic systems.

\begin{acknowledgments}
We thank T. Van\v{c}ura for his contribution to the experimental setup. Financial support from the Swiss Science Foundation (Schweizerischer Nationalfonds) and the NSF-NIRT is gratefully acknowledged. 
\end{acknowledgments}

\bibliography{apssamp}

\begin{references}

\bibitem{Grabert92} H.~Grabert and M.H.~Devoret, Eds., Single-charge tunneling (Plenum, New York, 1992).
\bibitem{Kouwenhoven97} L.P.~Kouwenhoven \textit{et al.}, in \textit{Mesoscopic Electron Transport}, L.P. Kouwenhoven, G. Sch\"on, and L.L. Sohn, Eds. (Kluwer, Dordrecht, Netherlands, 1997).
\bibitem{Reimann02} S.M.~Reimann and M.~Manninen, Rev. Mod. Phys. {\bf 74}, 1283 (2002).
\bibitem{Loss98} D.~Loss and D.P.~DiVincenzo, Phys. Rev. A {\bf 57}, 120 (1998).
\bibitem{Tarucha00} S.~Tarucha \textit{et al.}. Phys. Rev. Lett. {\bf 84}, 2485 (2000).
\bibitem{Luscher01} S.~L\"uscher \textit{et al.}, Phys. Rev. Lett. {\bf 86}, 2118 (2001).
\bibitem{Rokhinson01} L.P.~Rokhinson, L.J.~Guo, S.Y.~Chou, and D.C.~Tsui, Phys. Rev. B {\bf 63}, 035321 (2001).
\bibitem{Fuhrer03} A.~Fuhrer \textit{et al.}, Phys. Rev. Lett. {\bf 91}, 206802 (2003).
\bibitem{Fuhrer01} A.~Fuhrer \textit{et al.}, Nature {\bf 413}, 822 (2001).
\bibitem{Topinka01} M.A.~Topinka \textit{et al.}, Nature {\bf 410}, 183 (2001).
\bibitem{Topinka03} M.A.~Topinka, R.M.~Westervelt, and E.J.~Heller, Phys. Today {\bf 56}, 47 (2003).
\bibitem{Crook00} R.~Crook, C.G.~Smith, M.Y.~Simmons, and D.A.~Ritchie, J. Phys.: Condens. Matter {\bf 12}, L735 (2000).
\bibitem{Crook03} R.~Crook \textit{et al.}, Phys. Rev. Lett. {\bf 91}, 246803 (2003).
\bibitem{Yoo97} M.J.~Yoo \textit{et al.}, Science {\bf 276}, 579 (1997).
\bibitem{Yacoby99} A.~Yacoby \textit{et al.}, Solid State Commun. {\bf 111}, 1 (1999).
\bibitem{Zhitenev00} N.B.~Zhitenev \textit{et al.}, Nature {\bf 404}, 473 (2000).
\bibitem{Ilani04} S.~Ilani \textit{et al.}, Nature {\bf 427}, 328 (2004).
\bibitem{Finkelstein00} G.~Finkelstein, P.I.~Glicofridis, R.C.~Ashoori, and M.~Shayeagan, Science {\bf 289}, 90 (2000).
\bibitem{Woodside02} M.T.~Woodside and P.L.~McEuen, Science {\bf 296}, 1098 (2002).
\bibitem{Fuhrer02} A.~Fuhrer \textit{et al.}, Superlattices and Microstructures {\bf 31}, 19 (2002).
\bibitem{Ihn04} T.~Ihn, \textit{Electronic quantum transport in mesoscopic semiconductor structures}, Springer Tracts in Modern Physics 192 (Springer, New York, 2004).
\bibitem{Weichselbaum03} Simulations were carried out using the algorithm described in: A.~Weichselbaum and S.E.~Ulloa, Phys. Rev. E {\bf 68}, 056707 (2003).
\bibitem{Vancura03} T.~Van\v{c}ura \textit{et al.}, Appl. Phys. Lett. {\bf 83}, 2602 (2003).
\bibitem{Weichselbaum04} A.~Weichselbaum and S.E.~Ulloa, unpublished.
\end{references}

\end{document}